# Quantum Mechanics without an Equation of Motion

A. D. Alhaidari

*Saudi Center for Theoretical Physics, Jeddah, Saudi Arabia*

We propose a formulation of quantum mechanics in three dimensions with spherical symmetry for a finite level system whose dynamics is not governed by a differential equation of motion. The wavefunction is written as an infinite sum in a complete set of square integrable functions. Interaction in the theory is introduced in function space by a real finite tridiagonal symmetric matrix. Information about the structure and dynamics of the system is contained in the scattering matrix, which is defined in the usual way.

To study the structure and dynamics of a subatomic system, physicists perform scattering experiments. In such an experiment, a uniform flux of particles (referred to as projectiles or probes) will be impacted upon the target under study. The scattered flux carries the needed information about the system. In potential scattering theories [1], a potential function that models the structure of the system and its interaction with the probes is proposed. This model will be tested against the outcome of the experiment. Typically, such a potential function has a finite range such that in the asymptotic region, where the incident beam is ejected and the scattered particles are collected, its value is zero. Therefore, the incident and scattered particles are free and, thus, represented by sinusoidal wavefunctions. The information about the system (structure and dynamics) is contained in the difference between the two phase angles ("phase shift") of these sinusoidal functions. Mathematically, this means that one needs two independent functions that are asymptotically sinusoidal such as $\sin(kx)$ and $\cos(kx + \delta)$, where $x$ is the relevant configuration space coordinate and $k$ is the wave number which is related to the incident beam energy as $k = \sqrt{2mE/\hbar^2}$. The phase shift $\delta$ depends not only on kinematic quantities, such as the energy and angular momentum, but also on the model potential parameters [1].

In the above formulation, the model potential function is added to the appropriate kinetic energy operator to form the total Hamiltonian of the system. The traditional quantization method associates with this Hamiltonian a second order differential equation commonly known as the wave equation or the equation of motion (e.g., the Schrödinger equation) [2]. It is extremely rare to find an exact solution of this equation for a model potential that has more than few parameters (two or three). In fact, all the known exactly solvable quantum mechanical problems (about a dozen) are for potential functions with one or two parameters [3]. Moreover, for most of them that are in 3D with spherical symmetry, one is forced to restrict to the S-wave solution (zero angular momentum). Consequently, in the traditional quantum mechanics it is almost impossible to obtain *analytic* results associated with complex systems whose configuration requires many parameters to be described.

In this work, we propose an alternative formulation of quantum mechanics for systems with a finite structure (finite energy spectrum). Interaction is introduced without the need for a potential function, Hamiltonian operator, or an equation of motion. In the development, we employ some of the kinematic tools of the J-matrix method, widely used in atomic and nuclear scattering [4]. We expand the two wave functions representing the incident and scattered particles in terms of a complete set of square integrable functions. The expansion coefficients of these free wavefunctions satisfy a three-term recursion relation. On the other hand, the system's total wave function that includes the dynamics is also expanded in the same basis *but* with different expansion coefficients. Beside the kinematic parameters, these new coefficients depend also on the interaction that the system is subjected to. The elements of this interaction are included as real parameters in a modified three-term recursion relation to be satisfied by the new expansion coefficients. The asymptotic form of these coefficients will differ from those of the free wavefunction by a phase factor, which is in fact, the scattering matrix $e^{2i\delta(E)}$. As an illustration of the utility of the formulation, we give analytic formulas for the scattering matrix corresponding to problems with up to five interaction parameters. All the structural and dynamical information about the system is contained in this scattering matrix. In the following, we give brief technical details of the development and present some analytic, numerical, and graphical results.

For a given angular momentum quantum number $\ell$ and energy $E$ (whether for bound or continuum states), we assume that the wavefunction in three dimensions with spherical symmetry could be described faithfully by an infinite sum of a complete square integrable basis functions $\{\phi_n^\ell(r)\}_{n=0}^\infty$. That is, we can write $\psi_\ell(r,E) = \sum_n f_n^\ell(E)\phi_n^\ell(r)$, where the elements of the basis are chosen as follows

$$\phi_n^\ell(r) = a_n^\ell (\lambda r)^{\ell+1} e^{-\lambda^2 r^2/2} L_n^{\ell+\frac{1}{2}}(\lambda^2 r^2), \qquad r \geq 0. \tag{1}$$

$L_n^{\ell+\frac{1}{2}}(x)$ is the associated Laguerre polynomials [5] and $\lambda$ is a positive length scale parameter. The normalization constant is $a_n^\ell = \sqrt{2\Gamma(n+1)/\Gamma(n+\ell+3/2)}$. Therefore, the structural and dynamical information about the system is contained in the expansion coefficients $\{f_n^\ell(E)\}$, which depends also on the type of interaction that the system is subjected to. In the absence of interaction, we choose kinematic expansion coefficients, $\{s_n^\ell(E)\}$, such that $\sum_n s_n^\ell(E)\phi_n^\ell(r) = \mathcal{S}_\ell(r,E)$, where $\lim_{r\to\infty} \mathcal{S}_\ell(r,E) = \sin(kr - \pi\ell/2)$. In this work, we specialize to the case where $\mathcal{S}_\ell(r,E) = (kr)j_\ell(kr)$, where $j_\ell(x)$ is the spherical Bessel function. This function represents a free particle in 3D and one can confirm that this choice satisfies the asymptotic limit. With the help of the orthogonality relation of the Laguerre polynomials and tables of integrals, we obtain

$$s_n^\ell(E) = \sqrt{\pi/2}(-1)^n a_n^\ell \mu^{\ell+1} e^{-\mu^2/2} L_n^{\ell+\frac{1}{2}}(\mu^2), \tag{2}$$

where $\mu = k/\lambda$. Using the recursion relation satisfied by the Laguerre polynomials [5], we can show that $\{s_n^\ell(E)\}$ satisfy the following symmetric three-term recursion relation

$$\mu^2 s_n^\ell = \left(2n+\ell+\tfrac{3}{2}\right)s_n^\ell + \sqrt{n\left(n+\ell+\tfrac{1}{2}\right)}s_{n-1}^\ell + \sqrt{(n+1)\left(n+\ell+\tfrac{3}{2}\right)}s_{n+1}^\ell, \; n \geq 1, \tag{3a}$$

and its initial relation (for $n = 0$)

$$\mu^2 s_0^\ell = \left(\ell+\tfrac{3}{2}\right)s_0^\ell + \sqrt{\ell+\tfrac{3}{2}}\,s_1^\ell. \tag{3b}$$

Moreover, using the differential equation for the Laguerre polynomials, one can show that these coefficients satisfy the following second order differential equation in the energy

$$\left[ x \frac{d^2}{dx^2} + \frac{1}{2}\frac{d}{dx} - \frac{\ell(\ell+1)}{4x} - \frac{1}{4}x + \frac{1}{2}\left(2n + \ell + \tfrac{3}{2}\right) \right] s_n^\ell(E) = 0, \tag{4}$$

where $x = \mu^2$. Now, we introduce physical measurement by following the standard convention of the scattering matrix and its associated phase shift. That is, to account for the structure of the system and its interaction with the probes (test free particles), we measure the phase difference between the incident flux and scattered flux of these probes. Thus, we need to define two independent plane waves representing the incident and scattered free particle beams. Technically, this means that along with $\mathcal{S}_\ell(r,E)$ we need an additional wavefunction, say $\mathcal{C}_\ell(r,E)$, such that $\lim_{r\to\infty} \mathcal{C}_\ell(r,E) = \cos(kr - \pi\ell/2)$. Therefore, in the absence of interaction the asymptotic phase difference between $\mathcal{S}_\ell(r,E)$ and $\mathcal{C}_\ell(r,E)$ is $\pi/2$. We write $\mathcal{C}_\ell(r,E) = \sum_n c_n^\ell(E) \phi_n^\ell(r)$, where the new coefficients $\{c_n^\ell(E)\}$ are required to satisfy the following conditions:

(1) Be an independent solution of the 2$^{nd}$ order energy differential equation (4).
(2) Satisfy the three-term recursion relation (3a); but not its initial relation (3b).
(3) The new initial relation that replaces (3b) is chosen such that $\lim_{r\to\infty} \mathcal{C}_\ell(r,E) = \cos(kr - \pi\ell/2)$ not $\sin(kr - \pi\ell/2)$.

These three conditions result in the following new coefficients [6]

$$c_n^\ell(E) = \frac{(-1)^n}{\sqrt{\pi}} \Gamma\left(\ell + \tfrac{1}{2}\right) a_n \mu^{-\ell} e^{-\mu^2/2} {}_1F_1\left(-n - \ell - \tfrac{1}{2}; -\ell + \tfrac{1}{2}; \mu^2\right), \tag{5}$$

where ${}_1F_1(a;c;z)$ is the confluent hyper-geometric function [5]. Using this explicit form, we can show that the initial recursion relation for these new coefficients that should replace (3b) is

$$\mu^2 c_0^\ell = \left(\ell + \tfrac{3}{2}\right) c_0^\ell + \sqrt{\ell + \tfrac{3}{2}}\, c_1^\ell - \sqrt{2}\left(\mu/s_0^\ell\right). \tag{6}$$

Writing the three-term recursion relations (3a) in matrix form as $J^\ell(E)|s^\ell\rangle = 0$, then we can rewrite relations (3a), (3b) and (6) collectively as

$$J_{n,n-1}^\ell p_{n-1,\ell}^\pm + J_{n,n}^\ell p_{n,\ell}^\pm + J_{n,n+1}^\ell p_{n+1,\ell}^\pm = 0, \quad n = 1,2,3,..., \tag{7a}$$

$$J_{00}^\ell p_{0,\ell}^\pm + J_{01}^\ell p_{1,\ell}^\pm = \pm i2\sqrt{2}\mu/\left(p_{0,\ell}^\pm - p_{0,\ell}^\mp\right), \tag{7b}$$

where $J_{n,n}^\ell = \left(2n + \ell + \tfrac{3}{2}\right) - \mu^2$, $J_{n,n-1}^\ell = \sqrt{n\left(n + \ell + \tfrac{1}{2}\right)}$, and $p_{n,\ell}^\pm(E) = c_n^\ell(E) \pm i s_n^\ell(E)$.

Now, if we define the kinematical wavefunctions $\chi_\ell^\pm(r,E) \equiv \mathcal{C}_\ell(r,E) \pm i \mathcal{S}_\ell(r,E)$, then the asymptotic form of the system's total wavefunction becomes

$$\lim_{r\to\infty} \psi_\ell(r,E) = e^{-i\delta^\ell(E)} \chi_\ell^-(r,E) + e^{i\delta^\ell(E)} \chi_\ell^+(r,E) \tag{8}$$

where $\delta^\ell$ is the energy-dependent phase shift that also depends on the interaction for a given $\ell$ and $e^{2i\delta^\ell}$ is the scattering matrix $\mathbb{S}^\ell$. It has been shown elsewhere [7] that if we write $\Phi(r,E) = \sum_n g_n(E) \phi_n(r)$, then $\lim_{r\to\infty} \Phi(r,E) = \lim_{N\gg 1} \sum_{n=N}^\infty g_n(E) \phi_n(r)$. Hence, if we write $\{f_n^\ell\}$ in terms of new expansion coefficients $\{\hat{s}_n^\ell, \hat{c}_n^\ell\}$, then this result together with the asymptotic relation (8) suggest that for all $n \geq N$, where $N$ is some large enough integer, we can write

$$f_n^\ell = \tfrac{1}{2}\left(\hat{p}_{n,\ell}^- + \hat{p}_{n,\ell}^+\right) = \hat{c}_n^\ell, \quad \text{and} \tag{9a}$$

$$\hat{p}_{n,\ell}^\pm = e^{\pm i\delta^\ell} p_{n,\ell}^\pm. \tag{9b}$$

Therefore, the new expansion coefficients, $\{\hat{s}_n^\ell, \hat{c}_n^\ell\}$, satisfy the same asymptotic three-term recursion relation (7a) and we can write

$$J_{n,n-1}^\ell \hat{p}_{n-1,\ell}^\pm + J_{n,n}^\ell \hat{p}_{n,\ell}^\pm + J_{n,n+1}^\ell \hat{p}_{n+1,\ell}^\pm = 0, \quad n \geq N. \tag{10a}$$

On the other hand, the associated initial relation should be modified from (7b) and must incorporate the interaction. In fact, it should be a set of $N$ relations complementing (10a) and we write it as follows

$$\begin{aligned}&J_{n,n-1}^\ell \hat{p}_{n-1,\ell}^\pm + J_{n,n}^\ell \hat{p}_{n,\ell}^\pm + J_{n,n+1}^\ell \hat{p}_{n+1,\ell}^\pm = \pm i 2\sqrt{2}\mu\big/\left(p_{0,\ell}^\pm - p_{0,\ell}^\mp\right)\delta_{n0} \\ &-\left(\Omega_{n,n-1}\hat{p}_{n-1,\ell}^\pm + \Omega_{n,n}\hat{p}_{n,\ell}^\pm + \Omega_{n,n+1}\hat{p}_{n+1,\ell}^\pm\right)\end{aligned}, \quad n = 0,1,..,N-1 \tag{10b}$$

where $\{\Omega_{nm}\}_{n,m=0}^{N-1}$ are the elements of an $N \times N$ real tridiagonal symmetric matrix. Hence, the matrix $\Omega$ represents the interaction in this theory. The new asymptotic relation (10a) is the same as the original relation (7a). However, (10b) represent a new initial relation with $\Omega$ being a seed matrix with $2N-1$ real parameters characterizing the interaction. Now, to recover the original recursion (7a) and its initial relation (7b), we propose the following $2N-1$ parameter transformation of the complex coefficients

$$\hat{p}_{n,\ell}^\pm = \rho_n^\ell e^{\pm i\sigma_n^\ell} p_{n,\ell}^\pm, \quad n = 0,1,..,N-2, \tag{11a}$$

$$\hat{p}_{n,\ell}^\pm = e^{\pm i\delta^\ell} p_{n,\ell}^\pm, \quad n \geq N-1, \tag{11b}$$

where $\{\rho_n^\ell\}$ and $\{\sigma_n^\ell\}$ are real constant parameters, and $\rho_n^\ell > 0$. Transformation (11b) is inspired by Eq. (9b). It represents the asymptotic behavior of the expansion coefficients of the system's total wavefunction giving the phase shift $\delta^\ell(E)$ that depends on the interaction matrix $\Omega$ and the length scale parameter $\lambda$. Substituting from (11a) and (11b) into (10b) and using the original recursion (7a) and its initial relation (7b), we arrive at the analytic expression for the scattering matrix $e^{2i\delta^\ell(E)}$. Due to the fact that all elements that enter in the scattering matrix are functions of the quantity $\mu$ then all physical results exhibit the following scaling with $\lambda$: $\lambda \to \lambda'$ is equivalent to $E \to E(\lambda'/\lambda)^2$. The analytic result of the formulation for an interaction matrix $\Omega$ of rank three ($N = 3$) is:

$$e^{2i\delta^\ell} = T_0^\ell e^{-2i\xi_\ell} + \frac{1-T_0^\ell}{R_1^\ell \Lambda^\ell}\left(J_{01}^\ell + \frac{J_{00}^\ell}{R_1^\ell}\right)\left[\Lambda^\ell\left(J_{00}^\ell + \Omega_{00}\right) + \left(J_{12}^\ell - \Omega_{22}R_2^\ell\right)\frac{J_{01}^\ell + \Omega_{01}}{J_{12}^\ell + \Omega_{12}}\right]^{-1}, \tag{12}$$

where $T_n^\ell = p_{n,\ell}^- / p_{n,\ell}^+$, $R_{n+1}^\ell = p_{n+1,\ell}^+ / p_{n,\ell}^+$, $\xi_\ell = \arg(R_1^\ell \Lambda^\ell)$ and,

$$\Lambda^\ell = \left(J_{01}^\ell + \Omega_{01}\right)^{-1}\left[\left(J_{01}^\ell/R_1^\ell\right) + J_{11}^\ell - \Omega_{12}R_2^\ell + \frac{J_{11}^\ell + \Omega_{11}}{J_{12}^\ell + \Omega_{12}}\left(\Omega_{22}R_2^\ell - J_{12}^\ell\right)\right]. \tag{13}$$

It is easy to verify that if the interaction matrix vanishes then $e^{2i\delta^\ell(E)} = 1$. Moreover, any scattering matrix for a given rank is obtained by restriction from that of a higher rank. For example, taking $\Omega_{12} = \Omega_{22} = 0$ reduces (12) to the scattering matrix for $N = 2$. Additionally, it is our prediction that for a given interaction matrix $\Omega$ of rank $N$, the maximum number of bound states (the size of the structure of the system) is $2N-1$.

In Figure 1, we plot $\left|1-\mathbb{S}^\ell(E)\right|$ for the case $N = 1$ and for several values of the interaction parameter, $\Omega_{00}$. The plot clearly shows a resonance whose location on the energy axis and sharpness increase with $\Omega_{00}$. Table 1 is a list of these resonances. The bound states energy spectrum could be obtained from the poles of the scattering matrix. That is, the bound state energy $\tilde{E}$ could be obtained from (12), when restricted to $N = 1$, as solution of the equation

$$J_{00}^\ell(\tilde{E}) + J_{01}^\ell(\tilde{E})R_1^\ell(\tilde{E}) = -\Omega_{00}. \tag{14}$$

Table 2, lists the bound state energy for several values of $\Omega_{00}$ and $\ell$. Table 3 show similar results for the case $N = 2$ where the interaction has three dimensionless parameters ($\Omega_{00}$, $\Omega_{01}$, and $\Omega_{11}$) in addition to $\lambda$. Figure 2 is a sample result for the case $N = 3$.

In conclusion, we presented a formulation of quantum mechanics for a finite level system without the need for defining a potential function, Hamiltonian operator, or for solving a differential equation of motion. Interaction in this theory is given in function space by a finite tridiagonal symmetric matrix whose elements are the interaction parameters. The system's wavefunction is written as an infinite sum of square integrable functions with expansion coefficients that satisfy a three-term recursion relation incorporating the interaction parameters. The structure and dynamics are contained in the scattering matrix, which is defined in the conventional manner. Analytic results are obtainable for a much larger number of system parameters compared to the traditional potential method. This assertion was demonstrated for up to five configuration parameters. We conjectured that the maximum size of the energy spectrum (dimension of the system structure) is $2N - 1$, where $N$ is the rank of the interaction matrix.

**TABLES CAPTIONS:**

**Table 1**: The resonance energies shown in Fig. 1 as a function of $\Omega_{00}$.

**Table 2**: The bound state energy for $N = 1$ and for several values of $\Omega_{00}$ and $\ell$.

**Table 3**: Bound state energies for $N = 2$ and for several values of $\ell$. The interaction matrix is taken as $\Omega = \begin{pmatrix} 3 & 2 \\ 2 & 1 \end{pmatrix}$.

**Table 1**

| $\Omega_{00}$ | $E/\lambda^2$ |
|---|---|
| 3.0 | 2.840800 |
| 4.0 | 3.505974 |
| 5.0 | 4.044858 |
| 6.0 | 4.537273 |
| 7.0 | 5.011428 |
| 8.0 | 5.480449 |
| 9.0 | 5.950631 |
| 10.0 | 6.424600 |

**Table 3**

| $\ell$ | $-E/\lambda^2$ |
|---|---|
| 0 | 0.2062083314 |
|   | 1.7447577960 |
| 1 | 0.2058919014 |
|   | 0.2588596716 |
|   | 1.7991175093 |
| 2 | 0.4986143884 |
|   | 1.7370971472 |
| 3 | 0.4353327719 |
|   | 0.4978693172 |
|   | 1.6248260853 |

**Table 2**

| $\ell$ | $\Omega_{00}$ | $-E/\lambda^2$ |
|---|---|---|
| 0 | −0.5 | 0.2765686329 |
| 1 | −5.0 | 0.2010013828 |
|   | −3.0 | 0.1537776596 |
|   | +3.0 | 0.3331712440 |
| 2 | −1.0 | 0.4970185749 |
|   | −2.0 | 0.3397721281 |
| 3 | −5.0 | 0.3201382780 |
|   | +3.0 | 0.6090374585 |
|   | +5.0 | 0.5737587682 |

**FIGURES CAPTIONS:**

**Fig. 1**: Plot of $\left|1-\mathbb{S}^\ell(E)\right|$ versus energy for $N = 1$, $\ell = 1$ and for several values of the dimensionless interaction parameter, $\Omega_{00}$. These values are {1,2,...,10} and the larger the value of $\Omega_{00}$, the sharper the resonance and the more shift to higher energies.

**Fig. 2**: Plot of the phase shift angle $\delta^\ell(E)$, in units of $\pi$, versus energy for $\ell = 2$ and $N = 3$. The five elements of the interaction matrix are: $\Omega_{00} = 3$, $\Omega_{11} = 1$, $\Omega_{22} = -2$, $\Omega_{01} = -2$, and $\Omega_{12} = 1$. The plot clearly indicates sharp resonance activity near $E = 3.25\lambda^2$.

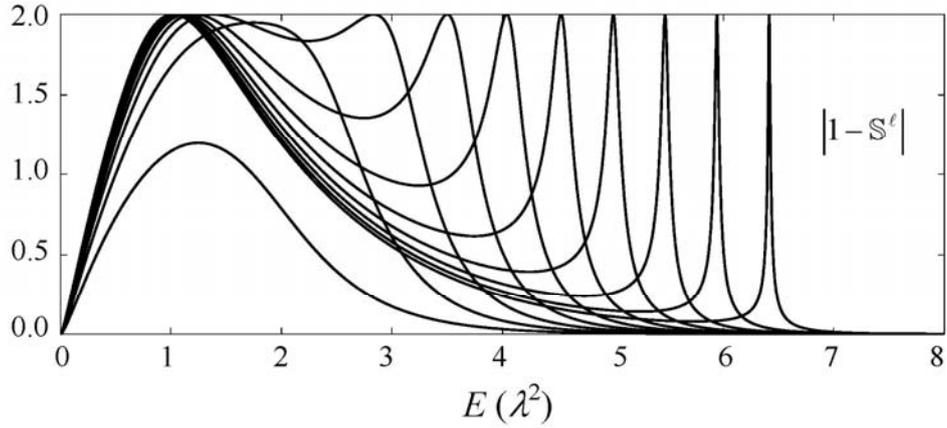

**Fig. 1**

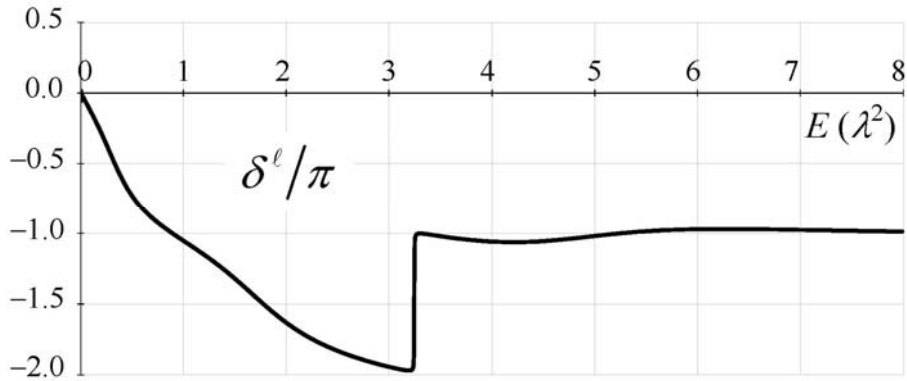

**Fig. 2**